# Aggregated knowledge from a small number of debates outperforms the wisdom of large crowds

Joaquin Navajas[1,2], Tamara Niella[1], Gerry Garbulsky[1], Bahador Bahrami[2,3] & Mariano Sigman[1]

[1]Universidad Torcuato Di Tella, Buenos Aires, Argentina

[2]Institute of Cognitive Neuroscience, University College London, United Kingdom

[3]Faculty of Psychology and Educational Sciences, Ludwig Maximilian University, Munich, Germany

**The aggregation of many independent estimates can outperform the most accurate individual judgment[1-3]. This centenarian finding[1, 2], popularly known as the "wisdom of crowds"[3], has been applied to problems ranging from the diagnosis of cancer[4] to financial forecasting[5]. It is widely believed that social influence undermines collective wisdom by reducing the diversity of opinions within the crowd. Here, we show that if a large crowd is structured in small independent groups, deliberation and social influence within groups improve the crowd's collective accuracy. We asked a live crowd (*N*=5180) to respond to general-knowledge questions (e.g., what is the height of the Eiffel Tower?). Participants first answered individually, then deliberated and made consensus decisions in groups of five, and finally provided revised individual estimates. We found that averaging consensus decisions was substantially more accurate than aggregating the initial independent opinions. Remarkably, combining as few as four consensus choices outperformed the wisdom of thousands of individuals.**

Understanding the conditions upon which humans benefit from collective decision-making has puzzled mankind since the origin of political thought[6]. Theoretically, aggregating the opinions of many unbiased and independent agents can outperform the best single judgment[1], which is why crowds are sometimes wiser than their individuals[2, 3]. This principle has been applied to many problems which include predicting national elections[7], reverse-engineering the smell of molecules[8], and boosting medical diagnoses[4]. The idea of wise crowds, however, is at odds with the pervasiveness of poor collective judgment[9]. Human crowds may fail for two reasons. First, human choices are frequently plagued with numerous systematic biases[10]. Second, opinions in a crowd are rarely independent. Social interactions often cause informational cascades, which correlate opinions, aligning and exaggerating the individual biases[11]. This imitative behaviour may lead to *herding*[9], a phenomenon thought to be the cause of financial bubbles[12], rich-get-richer dynamics[13, 14], and zealotry[15]. Empirical research has shown that even weak social influence can undermine the wisdom of crowds[16], and that collectives are less biased when their individuals resist peer influence[17]. Extensive evidence suggests that the key to collective intelligence is to protect the independence of opinions within a group.

However, in many of those previous works, social interaction was operationalized by participants observing others' choices without *discussing* them. These reductionist implementations of social influence may have left unexplored the contribution of deliberation in creating wise crowds. For example, allowing individuals to discuss their opinions in an online chat room results in more accurate estimates[18, 19]. Even in face-to-face interactions, human groups can communicate their uncertainty and make joint decisions that reflect the reliability of each group member[20, 21]. During peer discussion, people also exchange shareable arguments[22, 23], which promote the understanding of a problem[24]. Groups can reach consensuses that are outside the span of their individual decisions[24, 25], even if a minority[26] or no one[24] knew the correct answer before interaction. These findings lead to the following questions: can crowds be any wiser if they debated their choices? Should their members be kept as independent as possible and aggregate their uninfluenced, individual opinions? We addressed these questions by performing an experiment on a large live crowd (**Fig 1A,** see also **Supplementary Video 1**).

We asked a large crowd (N=5180, 2468 female, aged 30.1±11.6 years) attending a popular event to answer eight questions involving approximate estimates to general knowledge quantities (e.g., what is the height in meters of the Eiffel Tower? c.f. **Methods**).

Each participant was provided with pen and an answer sheet linked to their seat number. The event's speaker (author M.S.) conducted the crowd from the stage (**Fig. 1A**). In the first stage of the experiment, the speaker asked eight questions (**Supplementary Table 1**) and gave participants 20 seconds to respond to each of them (stage *i1*, left panel in **Fig. 1A**). Then, participants were instructed to organize into groups of five based on a numerical code in their answer sheet (see **Methods**). The speaker repeated four of the eight questions and gave each group one minute to reach a consensus (stage *c*, middle panel in **Fig. 1A**). Finally, the eight questions were presented again from stage and participants had 20 seconds to write down their individual estimate, which gave them a chance to revise their opinions and change their minds (stage *i2*, right panel in **Fig. 1A**). Participants also reported their confidence in their individual responses in a scale from 0 to 10.

Responses to different questions were distributed differently. To pool the data across questions, we used a non-parametric normalising method, used for rejecting outliers[27] (see **Methods**). Normalising allowed us to visualize the grouped data parsimoniously, but all our main findings are independent of this step (**Supplementary Fig. 1**). As expected, averaging the initial estimates from $n$ participants led to a significant decrease in collective error as $n$ increased ($F(4,999)= 477.3$, $p\sim 0$; blue lines in **Fig. 1B**), replicating the classic wisdom-of-crowd effects [2]. The average of all initial opinions in the auditorium (N=5180) led to 52% error reduction compared to the individual estimates (Wilcoxon sign rank test, $z=61.79$, $p\sim 0$).

We then focused on the effect of debate on the wisdom of crowds, and studied whether social interaction and peer discussion impaired[16, 17] or promoted[23, 24] collective wisdom. To disentangle these two main alternative hypotheses, we looked at the consensus estimates. We randomly sampled $m$ groups and compared the wisdom of $m$ consensus estimates (stage *c*) against the wisdom of $n$ initial opinions (stage *i1*, $n = 5\,m$ since there were 5 participants on each group). This analysis is based on the 280 groups (1400 participants) that had valid data from all of their members (see **Methods**). We observed that the average of as few as 3 collective estimates was more accurate than the mean of the 15 independent initial estimates (blue line at $n = 15$ vs. black line at $m = 3$ in **Fig. 1B**, $z$=13.25, $p$=10$^{-40}$). The effect was even more clear when comparing 4 collective choices against the 20 individual decisions comprising the same 4 groups (blue line at $n = 20$ vs. black line at $m = 4$ in **Fig. 1B**, $z$=20.79, $p$=10$^{-96}$). Most notably, the average of 4 collective estimates was even more accurate (by 49.2% reduction in error) than the average of the 1400 initial individual estimates (blue data point at $n = 1400$ vs. black line at $m = 4$ in **Fig 1B**, $z$=13.92, $p$=10$^{-44}$). In principle, this could simply result from participants having a second chance to think about these questions, and providing more accurate individual estimates to the group discussion than the ones initially reported. However, our data rules out this possibility since one or two collective estimates were not better than 5 or 10 independent initial estimates respectively (z=1.02, p=0.31). In other words, this is the result of a *crowd of crowds* (**Fig. 1C**).

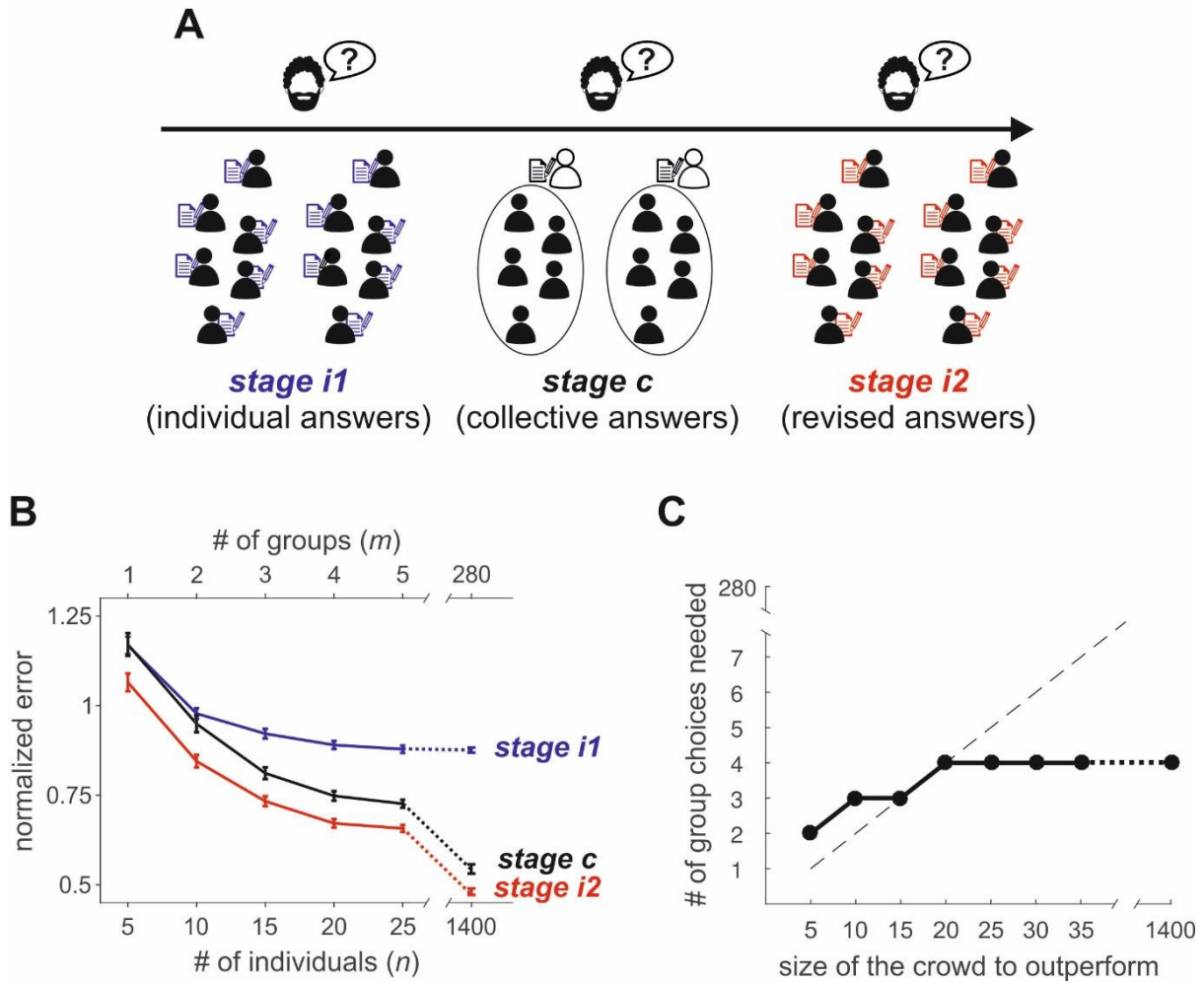

**Fig.1. Aggregating debates and the wisdom of crowds. (A)** A live crowd (N=5180) answered general knowledge questions in three stages. Left: initial individual estimate (stage *i1*). Middle: consensus (stage *c*). Right: revised individual estimate (stage *i2*). In stage *c*, a moderator (white) recorded the group's consensus estimate. **(B)** Normalized error of the average of $n$ individual answers (blue line for stage *i1*, red line for stage *i2*), and normalized error of the average of $m = n/5$ collective estimates (black line, stage *c*). Bars are s.e.m. **(C)** Minimum number of collective decisions needed to significantly (α = 0.01) outperform crowds of different sizes. At least 2 group estimates (from 10 individuals) are needed to outperform the wisdom of 5 independent individuals, and 3 estimates (from 15 individuals) are needed to outperform the wisdom of 10 independent estimates. For these crowd sizes, the wisdom of crowds is more efficient than aggregating debates. However, averaging 4 collective decisions leads to estimates that are significantly more accurate than the wisdom of crowds of any size.

Participants used the chance to change their minds after interaction and this reduced their individual error (mean error reduction of 31%, $z$=19.16, $p$=$10^{-82}$). More importantly, revised estimates gave rise to greater wisdom of crowds compared to initial estimates (blue line vs. red line in **Fig. 1B**, $F$(1,999)= 4458.6, $p$~0). When compared to collective choices, the average of $n$ revised decisions was overall more accurate than the average of $m$ group decisions (black line vs. red line in **Fig. 1B**, $F$(1,999)= 2510.4, $p$~0), although this depended on the specific question asked (interaction $F$(3,999)= 834.7, $p$~0; see **Supplementary Fig. 1**). Taken together, these findings demonstrate that face-to-face social interaction brings remarkable benefits in accuracy and efficiency to the wisdom of crowds. These results raise the question of how social interaction, which is expected to instigate herding, could have improved collective estimates.

Several observations about the bias and variance of the distributions of estimates help understanding our results (**Fig. 2**). **Fig. 2A** shows a graphical representation of how deliberation and social influence affected the distribution of responses in two exemplary groups. We found that the consensus decisions were less biased than the average of initial estimates (**Fig. 2B**, $z$=2.15, $p$=0.03, see also **Supplementary Fig. 2**). This indicates that deliberation led to a better consensus than what a simple averaging procedure (with uniform weights) could achieve. When participants changed their mind, they approached the (less biased) consensus: revised opinions became closer to the consensus than to the average of initial answers (**Fig. 2C**, $z$=27.15, $p$=$10^{-162}$). Moreover, in line with previous reports that social influence reduces the diversity of opinions[16, 17] we found that, *within each group*, revised responses converged towards each other: the variance of revised estimates within each group was smaller than the variance of the initial estimates (**Fig. 2D**, Wilcoxon sign rank test of the variance of responses on each group before vs after interaction, $z$=18.33, $p$=$10^{-75}$). However, interaction actually increased the variance of responses *between groups* (**Fig. 2E**): the distribution of the average of initial estimates (obtained by averaging stage *i1* estimates on each group) had less variance than the average of revised estimates (obtained by averaging stage *i2* estimates on each group, squared rank test for homogeneity, *p*<0.01). Previous research in social psychology also found a similar effect; consensus decisions are typically more extreme than the average individual choice, a phenomenon known as 'group polarization'[28].

Previous studies have proposed that a fundamental condition to elicit the wisdom-of-crowds effect is the diversity of opinions[3, 29]. Because we saw that interaction decreased the variance of estimates *within groups* but increased the variance *between groups*, we reasoned that sampling opinions from *different* groups might bring even larger benefits to the crowd. To test this idea, we sampled our population in two ways to test the impact of within- and between-group variance on the wisdom of crowds (**Fig. 2E**). In the *within-groups* condition, we sampled $n$ individuals coming from $m = n/5$ different groups. This was the same sampling procedure that we used in **Fig. 1B**. In the *between-groups* sampling, we selected $n$ individuals, each coming from a different group. Because different groups were randomly placed in different locations in the auditorium, we expected that sampling *between-groups* would break the effect of local correlations, and decrease the collective error.

Consistent with our predictions, we found that breaking the local correlations by between-group sampling led to a large error reduction (red solid line vs. red dashed line in **Fig. 2F**, 26% error reduction on average, $F(1,999)= 25824.1$, $p$~0). In fact, averaging only 5 revised estimates coming from 5 different groups outperformed the aggregation of all initial independent decisions in the auditorium ($z=25.91$, $p=10^{-148}$). This finding is consistent with previous studies showing that averaging approximately 5 members of "select crowds" leads to substantial increases in accuracy[30, 31]. In our case, adding more decisions using this sampling procedure led to a significant decrease in error ($F(4,999)= 249.34$, $p$~0). Aggregating revised estimates from different randomly sampled groups was a highly effective strategy to improve collective accuracy and efficiency, even with a very small number of samples.

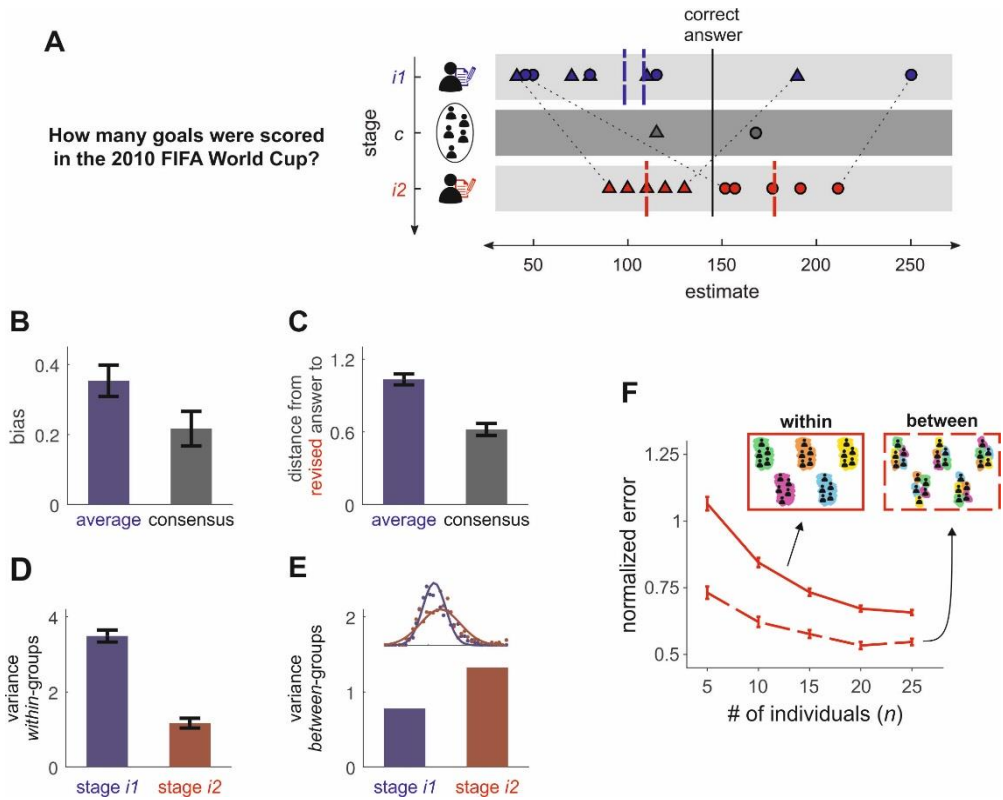

**Fig. 2. The effect of deliberation on bias and variance. (A)** Schematic illustration of the process of deliberation in two groups (circles or triangles) answering the question GOALS (see **Supplementary Table 1**) across the three stages of the experiment. Vertical dashed lines show the mean of each group. The range of opinions within each group decreased going from *i1* to *i2* (dotted black lines). The range of the *average* opinions (distance between vertical dashed lines) increased after deliberation. **(B)** Consensus decisions (grey bar) were less biased than the simple average of initial estimates in the group (blue bar). Bars show mean bias (signed error) and s.e.m. **(C)** Revised estimates (obtained in stage *i2*) were closer to the consensus decision (grey bar) than to the average of initial estimates (blue bar). Bars show mean distance and s.e.m. **(D)** Individuals conformed to the group consensus. Deliberation decreased the diversity of opinions within groups. Bars show variance within-groups (mean ± s.e.m.) before (blue bar) and after (red bar) deliberation. **(E)** Deliberation led to polarization of opinion and pulled groups to wider extremes in the opinion space. This process increased the diversity of opinions between different groups. The between-groups variance is obtained by taking the mean estimate on each group and computing the variance of this distribution across groups. The inset shows the distribution of mean estimates before (blue) and after (red) deliberation. Bars show variance between-groups. **(F)** We aggregated the individual estimates in two different ways: either by sampling participants all from the same groups (*within-groups condition*) or by sampling each participant from a different interacting group (*between-groups condition*). The insets sketch these two conditions; participants shaded by the same color were averaged together. The y-axis shows the normalised error of the average of $n$ individual answers at stage *i2* for the within-groups condition (solid line) and the between-groups condition (dashed line). Sampling participants who interacted in different debates leads to more accurate estimates.

We then asked whether deliberation was necessary to observe an increase in the wisdom of crowds. One could argue that the difference between wisdom of crowds obtained by aggregating the first (*i1*) versus the second (*i2*) opinions may have simply resulted from having a second chance to produce an estimate. Indeed, previous research[32-34] has shown consistent improvements drawn from repeatedly considering the same problem in decision making. To evaluate this possibility, we compared wisdom of crowds obtained from the answers to the discussed versus the undiscussed questions (see **Methods**). **Fig. 3A** shows the error reduction when comparing the average of $n$ revised estimates (*i2*) to the average of $n$ initial estimates (*i1*) i.e., the ratio of red line to blue line in **Fig 1B**. We observed that the error reduction in the absence of deliberation (**Fig 3A**, grey line) was below 3% for all crowd sizes. With deliberation (**Fig 3A**, green line), on the other hand, error reduction was significantly larger and increased with increasing the number of aggregated opinions ($F(1,999)$= 3963.6, $p$~0, comparing with- vs without deliberation). This result demonstrated that merely having the chance to produce a second estimate was not sufficient, and that deliberation was needed to the increase the wisdom of crowds.

While we found that deliberation increased collective accuracy, the results presented so far do not shed light on the specific deliberative procedure implemented by our crowd. In principle, collective estimates could have been the output of a simple aggregation rule different to the mean[17, 35]. Alternatively, participants could have used the deliberative stage to share arguments, and arrive to a new collective estimate through reasoning[22, 23]. This dichotomy between 'aggregating numbers' versus 'sharing reasons' has been discussed in several studies about collective intelligence[36, 37]. It has been argued that the normative strategy in predictive tasks is to share and aggregate numbers. Instead, problem-solving contexts require authentic deliberation and sharing of arguments and reasons[37]. Which kind of deliberative procedure did the groups implement in our experiment?

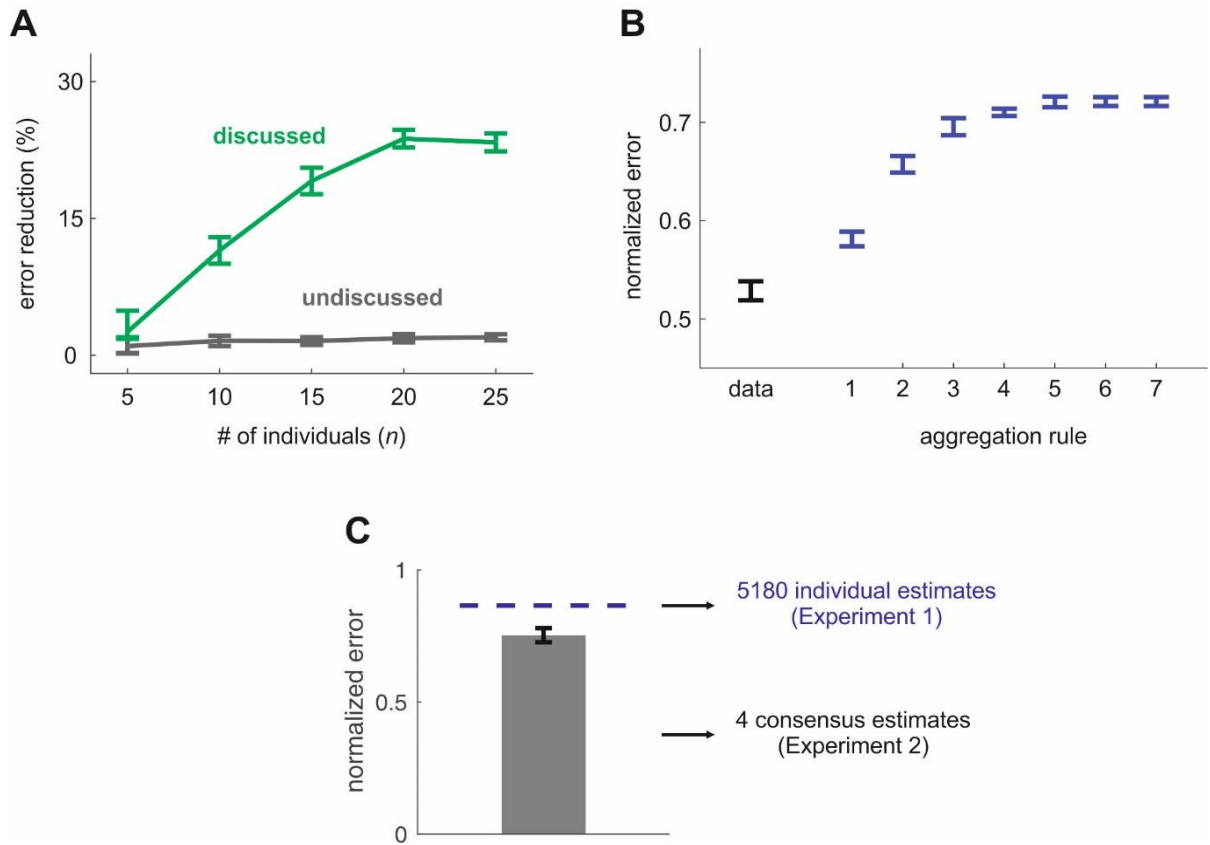

**Fig. 3. The superior wisdom of deliberative crowds. (A)** Collective accuracy in the absence of deliberation. Error reduction between averaging $n$ revised estimates compared to averaging $n$ initial estimates, expressed as percentual decrease. The green line shows the mean error reduction in the discussed questions and the grey line shows the same for the questions that remains undiscussed. Bars depict s.e.m. **(B)** Simple aggregation rules fail to explain the accuracy of deliberative crowds. Normalized error of averaging 100 randomly chosen collective decisions (black bars) versus averaging simulated estimates produced by seven simple aggregation rules (blue bars). Aggregation rules: (1) resistance to social influence, (2) confidence-weighted average, (3) expert rule, (4) median, (5) soft median, (6) mean, (7) robust average. See **Methods** for details about these rules. **(C)** We tested if four consensus choices could consistently and reliably outperform the wisdom of crowds. Aggregating 4 consensus estimates collected in the lab (Experiment 2) was more accurate than aggregating all 5,180 individual estimates from the crowd (Experiment 1). The y-axis shows error in normalized units (**see Methods**) and the error bar depicts s.e.m.

| **Deliberation Procedure** | Question 1 | Question 2 | Question 3 | Question 4 |
|---|---|---|---|---|
| *"We shared arguments and reasoned together"* | 8.0 ± 0.2 (10) | 7.5 ± 0.2 (10) | 7.9 ± 0.2 (10) | 7.4 ± 0.3 (8) |
| *"We followed the individuals who verbally expressed higher confidence during the debate"* | 6.4 ± 0.3 (8) | 6.0 ± 0.3 (8) | 6.1 ± 0.3 (8) | 6.9 ± 0.3 (10) |
| *"We discarded the estimates that were most far away from the mean"* | 5.4 ± 0.3 (0) | 5.0 ± 0.4 (0) | 5.5 ± 0.4 (0) | 5.3 ± 0.4 (0) |
| *"We followed the individuals who had reported higher confidence in the initial stage"* | 5.2 ± 0.4 (0) | 4.6 ± 0.4 (0) | 4.6 ± 0.4 (0) | 5.1 ± 0.4 (0) |
| *"We averaged our estimates"* | 5.0 ± 0.4 (0) | 4.4 ± 0.3 (0) | 4.0 ± 0.3 (0) | 3.5 ± 0.4 (0) |
| *"We followed the individuals who were least willing to change their minds"* | 2.3 ± 0.3 (0) | 2.4 ± 0.3 (0) | 2.1 ± 0.3 (0) | 3.3 ± 0.3 (0) |

**Table 1. Deliberation procedures implemented during the debates, as reported by the participants.** In Experiment 2, we replicated our main findings and asked participants to report the extent to which different deliberation procedures contributed to reaching consensus. Participants used a Likert scale from 0 to 10 (see **Methods** for details). The table show the mean rating ± s.e.m. for each question and procedure. Between brackets, we show the mode of the distribution of ratings (see also **Supplementary Fig. 4**). Procedures were sorted by mean rating, but participants rated them in randomized order. Question 1: GOALS. Question 2: ALEGRIA. Question 3: ROULLETTE. Question 4: OIL BARREL. (See **Supplementary Table 1** for details).

To answer this question, we first compared the accuracy of our consensus estimates with 7 different aggregation rules for how to combine the initial estimates (see **Methods**). Three of these rules were based on the idea of robust averaging[38], namely that groups may underweight outlying estimates (i.e., the median rule, the soft median rule and the robust averaging rule, see **Methods** for details). Three other rules were inspired in previous studies showing that the one individual may dominate the discussion and exert greater influence in the collective decision[21] (i.e., the expert rule, the confidence-weighted average rule, and the resistance-to-social-influence rule, see **Methods** for details). As a benchmark, we also compared these rules with the simple average rule. **Fig. 3B** shows the expected error if our crowd implemented each of these rules (blue bars in **Fig. 3B**). The empirically obtained consensus estimates (black bar in **Fig. 3B**) were significantly more accurate than all 7 aggregation rules ($z>3.99$, $p<10^{-5}$ for all pairwise comparisons between the observed data and all simulated rules). The deliberation procedures implemented by our crowd could not be parsimoniously explained by the application of any of these simple rules.

The above analysis definitively rejects the more simplified models of consensus. But the evidence is not exhaustive and does not necessarily imply positive evidence for the hypothesis that our crowd shared arguments during deliberation. To directly test this hypothesis, we ran a second experiment in the lab (Experiment 2, N=100, see **Methods** and **Supplementary Fig. 3**). Groups of 5 people first went through the experimental procedure (**Fig. 1A**). After finishing the experiment, in a debriefing questionnaire, we asked them what deliberation procedure(s) they implemented during the debates. After the end of stage *i2* (c.f. **Fig. 1A**), all participants were asked to rate (in a Likert scale from 0 to 10) the extent to which different deliberation procedures contributed to reaching consensus (see **Table 1** and **Supplementary Fig. 4**). The procedure with highest endorsement was "We shared arguments and reasoned together" (mean rating across all questions: 7.7±0.2, mode rating: 10; $z>7.05$, $p<10^{-12}$ for all comparisons, **Table 1**). We gave participants the opportunity to endorse more than one procedure or even describe a different procedure not appearing in our list. This latter option was selected less than 5% of the times (4.2 ± 2.0 %). Overall, our control analyses and new experiment suggest that (i) without deliberation there is no substantial increase in collective accuracy (**Fig. 3A**), (ii) the most salient simple aggregation rules previously proposed in the literature did not explain our findings (**Fig. 3B**), and (iii) participants reported sharing arguments and reasoning together during deliberation (**Table 1**).

Experiment 2 also allowed us to probe the wisdom of deliberative crowds *by design.* Since the materials (questions) and procedures were identical between the two experiments, we could formally test whether aggregating the consensus estimates (stage *i2*) drawn from 4 groups of 5 people collected in the lab could predictably and consistently outperform the aggregate of all independent opinions (stage *i1*) in the crowd. We found that the average of four group estimates collected in Experiment 2 was significantly more accurate than the average of all 5,180 initial estimates collected from the crowd ($z=6.55$, $p<10^{-11}$, **Fig. 3C**). It is difficult to overstate the importance of these findings as they call for re-thinking the importance of the *deliberation structure* in joint decision-making processes. This study opens up clear avenues for optimising decision processes through reducing the number of required opinions to be aggregated.

Our results are in contrast with an extensive literature on herding[11] and dysfunctional group behavior[39], which exhorts us to remain as independent as possible. Instead, our findings are consistent with research in collaborative learning showing that "think-pair-share" strategies[40] and peer discussion[24] can increase the understanding of conceptual problems. However, these findings offer a key novel insight largely overlooked in the literature on aggregation of opinions: pooling together collective estimates made by independent, small groups that interacted within themselves increases the wisdom-of-crowds effect. The potential applications of this approach are numerous and range from improving structured communication methods that explicitly avoid face-to-face interactions[41], to the aggregation of political and economic forecasts[42] and the design of wiser public policies[43]. Our findings thus provide further support to the idea that combining statistics with behavioural interventions leads to better collective judgments[18]. While our aim was to study a real interacting crowd, face-to-face deliberation may not be needed to observe an increase in collective accuracy. In fact, previous research has shown that social influence in virtual chatrooms could also increase collective intelligence[19,20].

The first study on the wisdom of crowds was regarded as an empirical demonstration that democratic aggregation rules can be trustworthy and efficient[2]. Since then, attempts to increase collective wisdom have been based on the idea that some opinions have more merit than others and set out to find those more accurate opinions by pursuing some ideal non-uniform weighting algorithm[17, 31, 35]. For example, previous studies proposed to select 'surprisingly popular' minority answers[35] or to average the responses of 'select crowds' defined by higher expertise[31] or by resistance to social influence[17]. Although these methods lead to substantial improvements in performance, implementing simple majority rules may still be preferred for other reasons which may include sharing responsibility[44], promoting social inclusion[39], and avoiding elitism or inequality[45, 46]. Here, we showed that the wisdom of crowds can be increased by simple face-to-face discussion within groups coupled with between-group sampling. Our simple -yet powerful- idea is that pooling knowledge from individuals who participated in independent debates reduces collective error. Critically, this is achieved without compromising the democratic principle of '*one vote, one value*'[47]. This builds on the political notion of deliberative polls, as a practical mechanism to solve the conundrum between *equality* and *deliberation*, as more and more people's voices are asked to make a decision and massive deliberation becomes impractical[48, 49]. Here, we demonstrated that in questions of general knowledge, where it is easy to judge the correctness of the group choice and in the absence of strategic voting behaviour[50], aggregating consensus choices made in small groups increases the wisdom of crowds. This result supports political theories postulating that authentic deliberation, and not simply voting, can lead to better democratic decisions[51].

## Methods

*Context*

The experiment was performed during a TEDx event in Buenos Aires, Argentina (http://www.tedxriodelaplata.org/) on September 24, 2015. This was the third edition of an initiative called *TEDxperiments* (http://www.tedxriodelaplata.org/tedxperiments), aimed at constructing knowledge on human communication by performing behavioural experiments on large TEDx audiences. The first two editions studied the cost of interruptions on human interaction[52], and the use of a competition bias in a "zero-sum fallacy" game[53].

*Materials*

Research assistants handled one pen and one A4 paper to each participant. The A4 paper was folded on the long edge and had four pages. On page 1, participants were informed about their group number and their role in the group. The three stages of the experiment (**Fig. 1A**) could be completed in pages 2, 3, and 4, respectively. On page 4, participants could also complete information about their age and gender.

*Experimental procedure*

The speaker (author M.S.) announced that his section would consist in a behavioral experiment. Participants were informed that their participation was completely voluntary and they could simply choose not to participate or withdraw their participation at any time. A total of 5180 participants (2468 female, mean age 30.1 years, s.d.: 11.6 years) performed the experiment. All data were completely anonymous. This experimental procedure was approved by the ethics committee of CEMIC (Centro de Educación Médica e Investigaciones Clínicas Norberto Quirno). A video of the experiment (**Supplementary Video 1**) is available in https://youtu.be/ND2-qRPERfk.

*Stage i1: individual decisions*

The speaker announced that, in the first part of the experiment, participants would make individual decisions. Subjects answered eight general knowledge questions that involved the estimation of an uncertain number (e.g., what is the height in meters of the Eiffel Tower?). Each question (**Supplementary Table 1**) had one code (e.g. EIFFEL) and two boxes. Participants were instructed to fill the first box with their estimate, and the second box with their confidence in a scale from 0 to 10. Before the beginning of stage *i1*, the speaker completed one exemplary question in the screen, and then read the eight questions. Participants were given 20 seconds to answer each question.

*Stage c: collective decisions*

In the second part (stage *c*), we asked participants to make collective decisions. First, they were instructed to find other members in their group according to a numerical code found in page 1. Each group had six members, and all participants were seated next to each other in two consecutive rows. The speaker announced that there were two possible roles in the group: player or moderator. Each group had five players and one moderator. Each participant could find their assigned role in page 1 (e.g., "You are the moderator in group 765" or "You are a player in group 391"). Players were instructed to reach a consensus and report it to the moderator in a maximum of 60 seconds. Moderators were given verbal and written instructions to not participate nor intercede in the decisions made by the players. The role of the moderators was simply to write down the collective decisions made by the players in their group. Moderators were also instructed to write down an 'X' if there was lack of consensus among the group. Groups were asked to answer four of the eight questions from stage *i1* (see **Supplementary Table 1**). The speaker read the four questions again, and announced the moments in which time was over.

*Stage i2: Revised decisions*

Finally, participants were allowed to revise all of their individual decisions and confidence, including the ones that remained undiscussed. The speaker emphasized that this part was individual, and read all eight questions of stage *i1* again.

*Data collection and digitalization*

At the end of the talk, we collected the papers as participants exited the auditorium. Over the week following the event, five data-entry research assistants digitalized these data using a keyboard. We collected 5180 papers: 4232 players and 946 moderators. Many of these 946 potential groups had incomplete data due to at least one missing player; overall, we collected 280 complete groups. All data reported in **Fig. 1** is based on those 280 complete groups (1400 players). For the comparison between individual, collective, and revised estimates, we focus on the four questions answered at stage *c*.

*Non-parametric normalization*

The distributions of responses were spread around different values on each question (**Supplementary Figure 1**). To normalize these distributions, we used a non-parametric approach inspired in the outlier detection literature [27]. We calculated the deviance of each data point $x_i$ around the median, and normalized this value by the median absolute deviance

$$n_i = \frac{x_i - median(\boldsymbol{x})}{median(|\boldsymbol{x} - median(\boldsymbol{x})|)}, \qquad [1]$$

where $\boldsymbol{x}$ is the distribution of responses. This procedure could be regarded as a non-parametric z-scoring of the data.

The rationale for normalizing our data was twofold. First, we used this procedure to reject outliers in the distribution of responses. Following previous studies[27], we discarded all responses with that deviated from the median in more than 15 times the median absolute deviance. The second purpose of normalization was to average our results across different questions. This helps the visualization of our data, but our findings can be replicated on each question separately without any normalization (**Supplementary Fig. 1**).

*Data analysis*

To compute all our curves in **Fig. 1**, we subsampled our crowd in two different ways: either by choosing $n$ individuals that interacted in $m = n/5$ different groups (within-groups sampling) or by choosing $n$ individuals from $n$ different groups (between-groups sampling). All curves in **Fig. 1B** and the solid line in **Fig. 2E** were based on the within-groups sampling condition; the dashed line in **Fig. 2E** is from the between-groups sampling condition. For a fair comparison between conditions, we computed the errors using exactly the same subsamples in our crowd. For each value of $n$, we considered 1,000 iterations of this subsampling procedure.

In the case of $n = 5$, each iteration randomly selected 5 of our 280 complete groups (**Fig 2E** sketches one exemplary iteration). In the within-groups condition, we computed the crowd error of each of the 5 groups (the error of the average response in stages *i1* and *i2*, and the error of the collective response in stage *c*) respecting the identity of each group. Finally, we averaged the 5 crowd errors and stored their mean value as the within-groups error for this iteration. In the between-groups sampling, we combined responses from individuals coming from different groups. We computed the error for 1,000 random combinations contingent on the restriction that all individuals belonged to different groups. Finally, we averaged all crowd errors and stored this value as the between-groups error for this iteration.

The same procedure was extended for $n > 5$. We randomly selected $n$ of our 280 groups on each of our 1,000 iterations. In the within-groups condition, we selected all possible combinations of $n$ individuals coming from $m$ groups, and computed their crowd error. We averaged the crowd error for all possible combinations and stored this value as the within-groups error for this iteration. In the between-groups condition, we randomly selected 1,000 combinations of $n$ individuals coming from $n$ different groups, and computed their crowd error. We averaged all of these crowd errors and stored this value as the between-subjects error for this iteration.

All error bars in **Fig. 1** and **2** depict the normalised mean ± s.e.m. of the crowd error across iterations. Pairwise comparisons were performed through non-parametric paired tests (Wilcoxon sign rank tests). To test the general tendency that error decreases for larger crowds, we used two-way repeated-measures analysis of variance (*rm-ANOVA*) with factors question and crowd side $n$, and iteration as repeated measure.

*Aggregation Rules*

We evaluated if collective estimates could result from 7 simple aggregation rules (**Fig. 3B**). All of these rules predict that the collective estimate $j$ is constructed using a weighted average of the initial estimates $x_i$ with weights $w_i$,

$$j = \sum_{i=1}^{5} w_i \, x_i \, .  \qquad [2]$$

In **Fig. 3B,** the seven rules were sorted by accuracy. Rule 1 is an average weighted by resistance to social influence. This procedure simulates that, during deliberation, the group follows the individuals who were least willing to change their minds, presumably because they had better information[17]. Resistance to social influence was quantified as the inverse linear absolute distance between the initial ($x_i$) and revised ($r_i$) estimates. This quantity was used to compute the weights

$$w_i = \frac{\sum_j |r_j - x_j + \varepsilon|^{-1}}{|r_i - x_i + \varepsilon|}, \qquad [3]$$

where $\varepsilon$ is a constant to prevent divergence when $x_j = r_j$. We simulated this rule using different values of $\varepsilon$ ranging from 0.1 to 1000, and used the value with highest accuracy ($\varepsilon = 1$). In Rule 2 (the 'confidence-weighted average rule'), the group uses the initial confidence ratings as weights in the collective decision, $w_i = c_i / \sum_j c_j$. In Rule 3, which we call the 'expert rule', the group selects the estimate of the most confident individual in the group. This rule is defined by $w_i = 1$ for $i = \text{argmax}(\mathbf{c})$, and $w_k = 0$ for $k \neq i$, where $\mathbf{c}$ is a vector with the five initial confidence ratings in the group.

Rule 4 consists in simply taking the median of the initial estimates, which is equivalent to giving a weight $w_i = 1$ to the third largest estimate in the group, and $w_k = 0$ to all other estimates. Rule 5 is the simple mean, namely $w_i = 0.2$ for all $i$. Rule 6, which we call "soft median", is a rule that gives a weight of $w_i = 0.5$ to the third largest estimate, weights of $w_k = 0.25$ to the second and fourth largest estimates, and $w_l = 0$ to the smallest and largest estimates in the group. Finally, Rule 7 is a robust average: this rule gives a weight $w_i = 0$ to all estimates in the group that differ from the mean in more than $k$ orders of magnitude, and equal weights to all other estimates. We simulated this rule using different values for $k$ ranging from 1 to 10, and used the value with highest accuracy ($k = 4$).

*Experiment 2*

N=100 naïve participants (56 female, mean age: 19.9 years, s.d.: 1.3 years) volunteered to participate in our study. Participants were undergraduate students at Universidad Torcuato Di Tella, and were tested as 20 groups of 5. The instructions and procedures were identical to the main task described above. At the end of the experiment, all individuals completed a questionnaire about the deliberation procedure implemented during the task. We asked them to rate (in a Likert scale from 0 to 10) the extent to which different deliberation procedures contributed to reaching consensus for each question. They rated six different procedures (see **Table 1** and **Supplementary Fig. 4**), which appeared in randomized order. We also gave them the possibility to choose "Other" and describe that procedure.

# Supplementary Figures

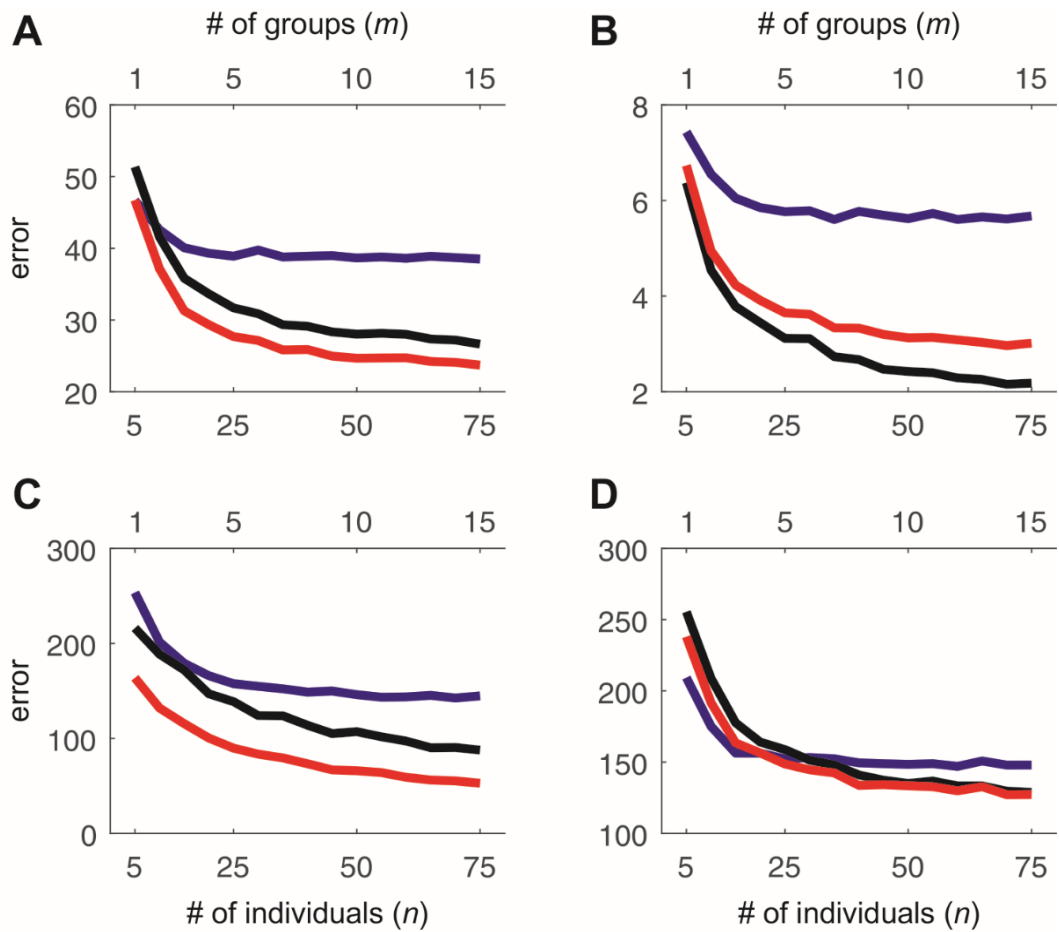

**Supplementary Figure 1. Error as a function of crowd size for each question**. Each panel show the same as **Fig. 1B** for each question separately and without normalization. Blue lines show the error obtained by averaging $n$ individual answers at stage *i1*, red lines show the same for stage *i2*, and black lines show the error of averaging $m$ collective estimates (black lines, stage *c*). All lines show the mean across 1,000 random within-group sub-samples of crowd; the s.e.m. is within the thickness of each line. **A)** GOALS: How many goals were scored in the 2010 FIFA World Cup? **B)** ALEGRIA: How many times does the word "alegría" (joy) appear in the lyrics of the song "Y dale alegría a mi corazón"? ("Give joy to my heart") **C)** ROULETTE: What is the sum of all numbers in a roulette wheel? **D)** OIL BARREL: How much did a barrel of oil cost in 1970 (in US dollars cents)?

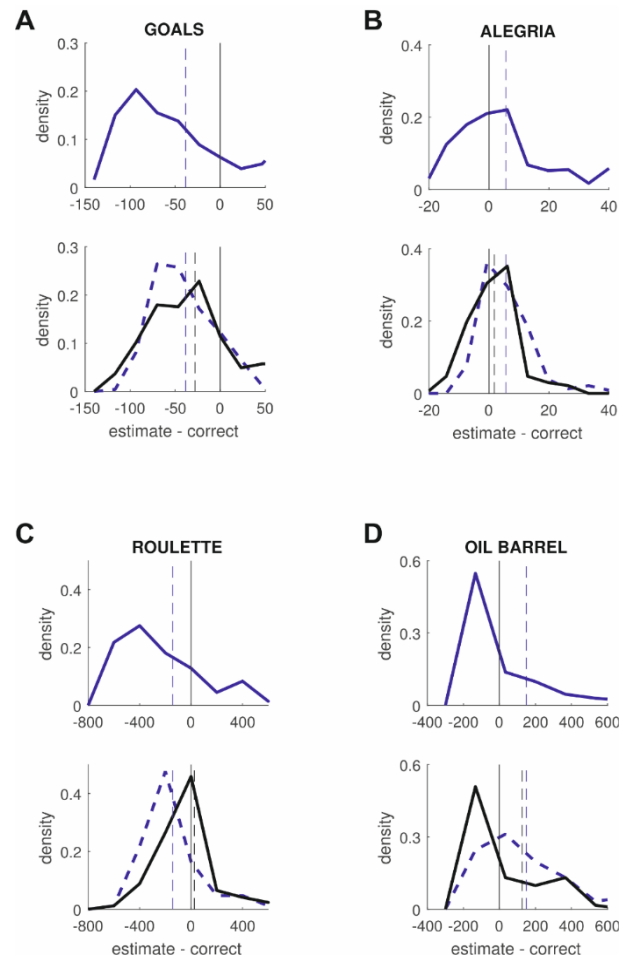

**Supplementary Figure 2. Distribution of responses relative to the correct answer**. On each question, the top panel is the normalized histogram of responses obtained at stage *i1* (blue solid line). The vertical dashed line is the mean of that distribution. The bottom panel show the distribution of mean estimates on each group (blue dashed line) and the distribution of consensus decisions of each group (black line). If the consensus reflected an average with uniform weights for each participant, then these two distributions should be identical. However, we observe that in the four questions, collective answers are less biased than the mean of all initial estimates of each group (black dashed line is closer to 0 than the vertical dashed line). This is why the wisdom of m collective answers is more accurate than the wisdom of n=m/5 initial estimates. Each panel **A-D** shows a different question. See **Supplementary Table 1** for more details.

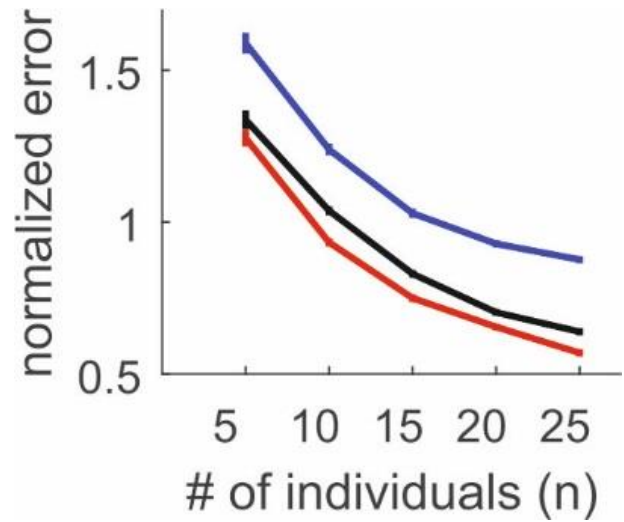

**Supplementary Figure 3. Replication in a lab experiment.** 20 groups of 5 individuals performed our experiment in a lab setting. The figure shows the normalized error across the four discussed questions for crowds with different number of individuals. The only difference between this figure and **Fig. 1b** is that it was produced with 100 (instead of 1,000) random subsamples of the population. This is because we have a sample size of 100 subjects compared to the 1,400 used in the main experiment. Blue line: average of $n$ initial estimates. Black line: average of $n/5$ collective estimates. Red line: average of $n$ revised estimates. The error bars denoting s.e.m. are within the thickness of the lines.

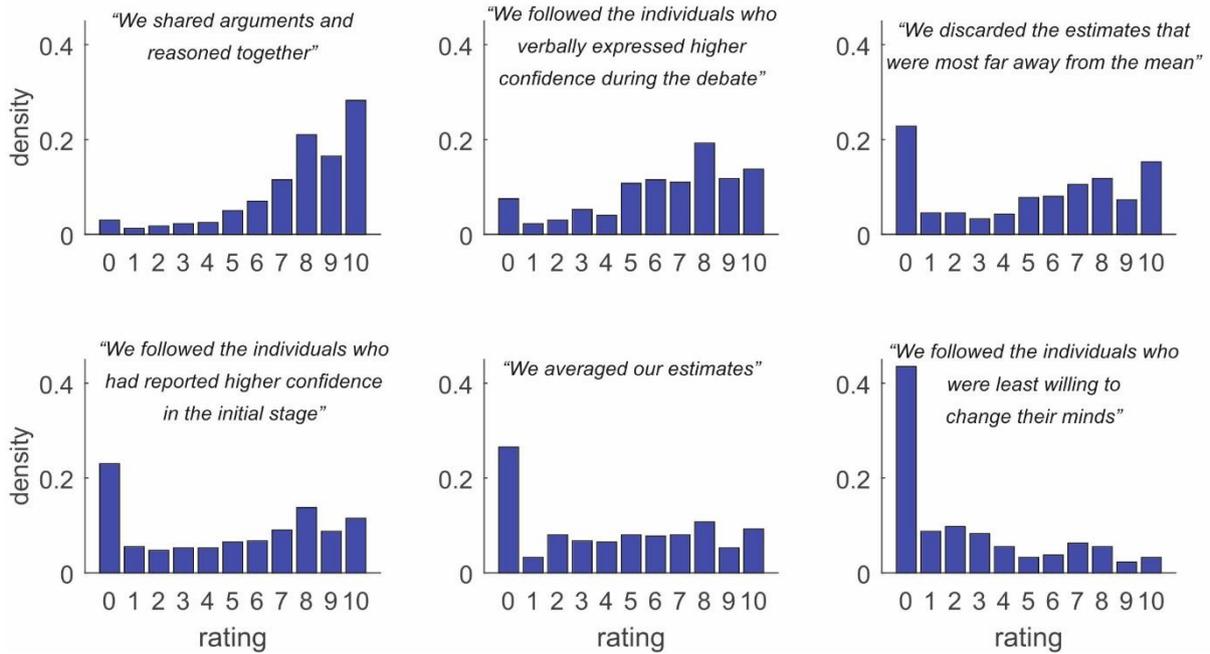

**Supplementary Figure 4. Distribution of ratings in the lab experiment.** For each deliberation procedure, bars show the distribution of ratings across all questions. As in **Table 1**, procedures were sorted by mean rating but participants observed them in randomized order. The procedure with highest rating was "We shared arguments and reasoned together" (7.7 ± 0.2). The second highest rating was for "We followed individuals who verbally expressed higher confidence during the debate" (6.3 ± 0.3). We found that this procedure (based on verbal communication of confidence) was perceived to contribute significantly more to the consensus than the procedure based on reported confidence ("We followed the individuals who had reported higher confidence in the initial stage", 4.9 ± 0.4). This is consistent with a previous study showing that aggregation rules based on confidence reports are suboptimal and hinder collective intelligence (Büchel, B., Klössner, S., Lochmüller, M., & Rauhut, H. 2017. The Strength of Weak Leaders–An Experiment on Social Influence and Social Learning in Teams. Working papers SES. Université de Friburg).

| Question | Code | Correct Answer | Mean | Median | Median Absolute Deviance | Discussed in groups? |
|---|---|---|---|---|---|---|
| How many goals were scored in the 2010 FIFA World Cup? | GOALS | 145 | 107.9 | 81 | 38 | Yes |
| How many emperors did the Roman Empire have? | EMPERORS | 131 | 19.5 | 10 | 5 | No |
| How many times does the word "alegría" (joy) appear in the lyrics of the song "Y dale alegría a mi corazón"? ("Give joy to my heart") | ALEGRIA | 21 | 26.7 | 21 | 9 | Yes |
| How many calories are there in 200 grams of butter? | BUTTER | 1480 | 1251.3 | 600 | 400 | No |
| What is the sum of all numbers in a roulette wheel? | ROULLETTE | 666 | 705.9 | 400 | 250 | Yes |
| How much did a barrel of oil cost in 1970 (in US dollars cents)? | OIL BARREL | 180 | 6445.4 | 150 | 138 | Yes |
| What is the height of the Eiffel tower in meters? | EIFFEL | 324 | 344.4 | 200 | 110 | No |
| How many elevators are there in the Empire State Building of Ney York? | ELEVATORS | 73 | 18.6 | 11 | 6 | No |

**Supplementary Table 1.** Questions, correct answers, and summary statistics of the distribution of responses before interaction (stage *i1*). From the eight questions asked in stage *i1*, half of them were discussed in groups (stage *c*). Distributions were normalized using a non-parametric method based on the median (column 5) and the median absolute deviance (column 6). See Equation [1] in **Methods** for details.

**Supplementary Video 1.** A video describing the experimental procedure and showing the crowd performing the experiment is available at https://youtu.be/ND2-qRPERfk.